\RequirePackage{lineno}
\documentclass[twocolumn,floatfix,superscriptaddress]{revtex4}

\usepackage{amsmath}
\usepackage{graphicx,xcolor}	

\newcommand{\eq}[1]{\begin{equation}{#1}\end{equation}} 

\newcommand{\figref}[1]{Fig.~\ref{#1}}		
\newcommand{\secref}[1]{Section~\ref{#1}}	
\newcommand{\tabref}[1]{Tab.~\ref{#1}}		
\newcommand{\appref}[1]{the online material}		

\newcommand{\diff}[2]{\frac{{\rm d} {#1}}{{\rm d{#2}}}} 
\newcommand{\vect}[1]{{#1}}			   
\newcommand{\matr}[1]{{\rm \bf {#1}}}		   
\newcommand{\ttabs}[1]{\lvert{#1}\rvert}	

\newcommand{\ttfrac}[2]{{#1}/{#2}} 

\begin{document}
 
\title{How to predict community responses to perturbations in the face of imperfect knowledge and network complexity}

\author{Helge \surname{Aufderheide}}
\affiliation{Max-Planck-Institute for the physics of complex systems, Dresden, Germany}
\affiliation{University of Bristol, Merchant Venturers School of Engineering, Bristol, UK}
\author{Lars \surname{Rudolf}}
\author{Thilo \surname{Gross}}
\affiliation{University of Bristol, Merchant Venturers School of Engineering, Bristol, UK}
\author{Kevin D. \surname{Lafferty}}
\affiliation{U.S. Geological Survey, Western Ecological Research Center, Marine Science Institute, University of California, Santa Barbara, CA 93106 }

\begin{abstract}
It is a challenge to predict the response of a large, complex system to a perturbation. Recent attempts to predict the behaviour of food webs have revealed that the effort needed to understand a system grows quickly with its complexity, because increasingly precise information on the elements of the system is required. Here, we show that not all elements of the system need to be measured equally well. This suggests that a more efficient allocation of effort to understand a complex systems is possible. We develop an iterative technique for determining an efficient measurement strategy. Finally, in our assessment of model food webs, we find that it is most important to precisely measure the mortality and predation rates of long-lived, generalist, top predators. Prioritizing the study of such species will make it easier to understand the response of complex food webs to perturbations.
\end{abstract}

\maketitle 
\section{Introduction\label{sec:Introduction}}
Predicting the result of environmental perturbations, such as the arrival of new species or habitat change, is a major goal in ecology \cite{sakai_population_2001,abrams_evolution_1996,parker_impact:_1999}. 
What makes this challenging is the complex interconnected nature of ecological systems.
In any densely-connected system, a perturbation of one element can percolate across the network of interactions.
This is particularly true for the complex food webs that form the backbones of most ecosystems 
\cite{montoya_press_2009,mooney_evolutionary_2001,mack_impacts_1998,pascual_ecological_2005}.
Even perturbations acting only on a small subset of species may thus propagate through the network and lead to serious systemic changes \cite{parker_impact:_1999,clavero_invasive_2005,mack_impacts_1998,menge_indirect_1995} 
such as the destruction of native fish populations following the introduction of a new species \cite{zaret_species_1973}.

A central factor determining the response of a food web to perturbations is its topology, the precise map of predator-prey interactions. It has been shown that topological properties affect local and global dynamical stability \cite{may_qualitative_1973,mccann_diversitystability_2000,martinez_artifacts_1991,gross_generalized_2009,dambacher_relevance_2002}. 
Often, the food web topology gives an indication of the relative importance of species when studying  notions of robustness such as the likelihood of secondary extinctions \cite{dunne_network_2002,jordan_keystone_2009,allesina_who_2004,libralato_method_2006}.

Topology alone, however, is generally not sufficient for reliable predictions of perturbation effects \cite{mills_keystone-species_1993,berlow_simple_2009,stouffer_compartmentalization_2011,montoya_press_2009}. Approaches taking into account biomass flows between the different species correlate much better with the results of numerical simulations than do topological indices alone \cite{jordan_identifying_2008}. Simulations have been successfully used to study species interaction strength \cite{berlow_simple_2009}, or topological compartmentalization \cite{stouffer_compartmentalization_2011} for perturbations that result from species extinctions. 

An accurate prediction of the impact of perturbations requires information about underlying biomass flows and the control coefficients characterizing the nonlinearity of processes. 
Such parameters require extensive measurements and 
errors in their estimation quickly reduce the accuracy of predictions  \cite{yodzis_indeterminacy_1988,montoya_press_2009}. 
A lack of precise information on biomass flows and control coefficients thus limits our ability to make precise predictions on the ultimate effect of perturbations. For instance, predictions for systems of more than $25$ species were found practically impossible with current methods, unless very detailed information is available \cite{novak_predicting_2011}. 

Here, we ask if food web responses to a perturbation are more sensitive to particular species or parameters \cite{klemm_measure_2012,jordan_keystone_2009,berg_using_2011}. If we can identify such influential elements in advance, we should be able to make more precise predictions about dynamics with a given effort in parametrization. 

In this paper, we investigate the predictability of responses to perturbations in a broad class of food web models.
Our results show that all parameters and species do not need to be measured with the same accuracy.
We use analytical calculations and numerical demonstrations to show that it is possible to assign to each species a value that indicates the importance of precise knowledge about this species for the quality of the prediction. 
Furthermore, we demonstrate that this importance can be estimated reasonably well from imprecise information. Finally, we identify which of the parameters of these important species are crucial to the prediction of responses to perturbations.

The paper is structured as follows: We start in \secref{sec:method} by introducing a method for predicting the impact of given perturbations in a broad class of food web models. The method is illustrated in \secref{sec:examples} with two examples. In \secref{sec:importance}, we then derive measures for species' influence on others and for their  sensitivity to perturbations. In \secref{sec:errors}, we test these predictions in a series of numerical experiments. The numerical results illustrate a feasible strategy for field studies, where mathematical analysis and experimental measurements are used to iteratively improve predictions about the response of food webs to disturbance.  In \secref{sec:statistical}, we use computer experiments and statistical association to determine which parameters and types of species are most important to measure.

\section{Impact Evaluation\label{sec:method}}

\subsection{Derivation of the perturbation impact}
Consider a biological system described by a set of state variables $X_1,\ldots,X_N$  denoting, for instance, the abundances of established species in a food web. The system is now subject to a perturbation that is characterized by another set of variables $Y_1,\ldots,Y_M$, for instance denoting the abundances of newly arriving species. 

We assume that, in the absence of the perturbation, the variables $X_1,\ldots,X_N$ are governed by a set of ordinary differential equations of the form
\begin{equation}
\label{eq:ODE}
\frac{\rm d}{\rm dt} X_i = A_i(X_1,\ldots,X_N),
\end{equation}
where $A_i$ is a function representing the \emph{right-hand-side} of the differential equations.
For instance, the generalized model for food webs \cite{gross_generalized_2006} which we use below, describes the dynamics of the populations $X_1,\ldots,X_N$ by $N$ differential equation of the form
\begin{equation}
\diff{}{t}X_i  = G_i(\vect{X}) + S_i(X_i) - L_i(\vect{X}) -M_i(X_i),
\end{equation}
where $G_i$, $L_i$, $M_i$, and $S_i$ are unspecified functions describing, respectively, the gain by predation ($G_i$), the loss by predation ($L_i$), the loss due to natural mortality ($M_i$), and the gain by primary production ($S_i$) of the focal species.

Following Ref.~\onlinecite{novak_predicting_2011}, we consider the case where the unperturbed system resides in a stable equilibrium $\vect{X}^*$ and where  the perturbation is characterized by a small and constant $\vect{Y}^*$, such as a new species persisting at a low constant abundance in the ecosystem due to initially positive growth or constant influx. This new species affects the right-hand-side of \eqref{eq:ODE}, i.e.  $A_i(\vect{X},\vect{Y})$ .    
 
Because the stationary abundance, $X_i^*$, of a given established species $i$ is dependent on the new species $\vect{Y}^*$, we can regard it as a function
$X_i^*=X_i^*({\vect{ Y}^*})$.
We can then define the impact $I_{i,j}$ of a perturbation variable $Y^*_j$ on a resident species abundance $X^*_i$ as the change of $X_i^*$ per unit $Y^*_j$, i.e.
\begin{equation}
\label{eq:IDef}
I_{i,j} = \left. \frac{\partial X_i^*}{\partial Y^*_j} \right|_0,
\end{equation}
where we used $|_0$ to indicate that the derivative is evaluated in the limit of vanishing densities of the arriving species $Y^*_j$.
In other words, the entries of the impact matrix $I_{i,j}$ state the loss of units of the established species $i$ per unit of arriving species $j$ that enters the system. 

For two to three species, the impact can be computed by first defining the model functions, then solving \eqref{eq:ODE} for the stationary solution $X_i^*({\vect{Y}^*})$, and subsequently computing the derivative in \eqref{eq:IDef}. However, for more than three species, the analytical computation of the stationary solution becomes prohibitively difficult. Furthermore, we seek a general solution, that does not depend on the functional forms  in the model.
For these reasons, the explicit computation of the stationary solution is not possible. 

Computing the stationary solution can be avoided by recognizing that the stationary density of a resident species $X_i^*$ can be considered as an implicit function that is defined as the solution of the stationarity condition 
$0=A_i(\vect{X}^*,\vect{Y}^*)$.
Using a corollary to the implicit function theorem \cite{khalil_nonlinear_1992}, we can then write the impact matrix as
\begin{equation}
\label{eq:ImpactMatrix}
{\matr{I}} = -{\matr{J}}^{-1} {\matr{K}},
\end{equation}
where the superscript $-1$ indicates the matrix inverse. 
The matrix ${\matr{J}}$ is the so-called Jacobian, which is defined as the derivatives of $A_i$ with respect to the abundances of established species \cite{kuznetsov_elements_1998}, i.e.
$J_{i,j}=\left.\ttfrac{\partial A_i}{\partial X_j}\right|_{*}$,
where $|_*$ indicates that the derivative is evaluated in the equilibrium under consideration.
And, the matrix ${\matr{K}}$ is a $N \times M$ defined by $K_{i,j}=\left.\ttfrac{\partial A_i}{\partial Y_j} \right|_{*}$.
The matrix ${\matr{K}}$ thus captures the direct impact of an arriving species on an established species. For instance, this direct impact may occur due to a reduction in production, an increase in mortality, or an increase in predation gain of the established species per unit of the arriving species. To establish ${\matr{K}}$ prior to the arrival therefore requires information about the resident species with whom the arriving species is likely to interact.

We note that, beyond the example of perturbations caused by an arriving species, \eqref{eq:ImpactMatrix} applies to press perturbations on an established community in general. 
For instance, to estimate the ultimate impact of a slight drought, the entry $K_{i}$ of the perturbation vector indicates the direct impact of the drought on species $i$; mathematically, we can approximate $K_i=\Delta A_i(\vect{X}^*)$, where $\Delta A_i(\vect{X}^*)$ denotes the absolute change of $A_i$ due to the drought.

In summary, \eqref{eq:ImpactMatrix} establishes a relationship between the direct proximal impact of a press perturbation, and the indirect ultimate impact ${\matr{I}}$, when the internal interactions among in the unperturbed system ${\matr{J}}$ are known.  As a note of caution, we remark that \eqref{eq:ImpactMatrix} relationship holds up to linear order. The impact-approximation therefore remains valid only as long as the perturbation caused is reasonably small.

\subsection{Parametrization of $\matr{J}$}
In the equations above, we refer to the steady state of the system, which seems to imply that information about this state is required.
However, relationship \eqref{eq:ImpactMatrix} remains valid independently of the specific steady state under consideration.
When the matrices are evaluated, the steady state appears only in the Jacobian which contains elements of the form $J_{i,j}=\ttfrac{\partial(\ttfrac{{\rm d} X_i}{{\rm dt}} )}{\partial X_j}|_*$.
For instance, in the generalized food web model (see the online material for a  review), this leads to expressions such as $\left.\ttfrac{\partial G_i}{\partial X_i}\right|_*$ \cite{gross_generalized_2006}.
Because we cannot evaluate this expression without further assumptions, it is an unknown quantity. However, we note that for any specific system the expression is simply a number.  
In other words, this means that the unknown derivatives appearing in the Jacobian constitute unknown parameters of the model.

So far, we have recognized that the unknown derivatives can be formally treated as unknown parameters. 
However, as such, these parameters are hard to interpret and are thus not suitable for an ecological discussion of the results. We solve this problem by using a slightly different parametrization, which is obtained either by a special normalization procedure \cite{gross_generalized_2006} or directly by the identity
\begin{equation}
\label{eq:Identity}
\left.\frac{\partial G_i}{\partial X_i}\right|_*=\frac{G_i^*}{X_i^*} \left.\frac{\partial \log G_i}{\partial \log X_i}\right|_*,
\end{equation}
which is true for $G_i^*,X_i^*>0$ (a condition that is generally met by definition; see Ref.~\onlinecite{kuehn_dynamical_2012} for the special case of $X_i^*=0$).

The expression on the right-hand-side of \eqref{eq:Identity} is a product of two factors that have a direct interpretation in most applications.
The first factor is a per-capita rate.
Such rates have the dimension of inverse time and can be directly interpreted as characteristic turnover rates, in this case, as the per-capita growth rate of the members of species $i$ by predation on other species. 

The second factor in \eqref{eq:Identity} is a logarithmic derivative.
Such derivatives are also called elasticities and have been proposed originally in economic theory \cite{houthakker_income_1969} and subsequently in metabolic control theory \cite{fell_metabolic_1992} and ecology \cite{yeakel_generalized_2011}. 
They can be estimated well from observational data and interpreted straightforwardly.
For every power-law, $f(x)=Ax^p$, the logarithmic derivative is $\ttfrac{\partial \log f}{\partial \log x} = p$, independently of $A$ or $x$.
Thus, for instance, any linear function has an elasticity of one regardless of the slope.
For functions that are not power-laws the elasticity still provides an intuitive non-linear measure of the sensitivity in the steady state.

\begin{table}
    \scriptsize
     \begin{tabular}{lll}
      \def\tlsp{1pt}
     \textbf{Name} & \textbf{Interpretation}  & \textbf{Value(s) } \\[2pt]
     \hline \\[-6pt]
\multicolumn{3}{l}{\textbf{Scale parameters} - defining the biomass flows in the steady state.}\\
$\alpha_i$ & Rate of biomass turnover in species $i$ &  $\alpha_i=0.08^{n_i}$ \\[0pt]   
$\beta_{i,j}$ & Contribution of predation by $i$     	& 0 (no feeding)	 \\[0pt]
              & to loss rate of species $j$ 	      	& [0.1,1] (else)\\
$\chi_{i,j}$ & Contribution of species $i$ 		& 0 (no feeding)\\[0pt]
             & to the prey of species $j$ 		&[0.1,1] (else)		\\
$\rho_i$     & Fraction of growth in species $i$     	& 0 (producers)	\\[0pt]
                    &    gained by predation, not primary production    	& 1 (consumers) \\
$\sigma_i$ & Fraction of mortality in species $i$    & 0 (top pred.)  	\\[0pt]
           &    resulting from predation, not mortality             & [0.5,1] (others) 	\\
\multicolumn{3}{l}{\textbf{Elasticities} - Sensitivities of interactions to state variables.}\\
$\gamma_i$ & Sensitivity of predation in species $i$ & [0.5,1.5]       	\\[0pt]
                     & to $i$'s prey density     	     & 			\\
$\lambda_{i,j}$ & Exponent of prey switching & 1 (passive) 		\\
$\mu_i$ & Exponent of closure in species $i$&  [1,2]        		\\
$\phi_i$ & Sensitivity of primary production in species $i$  & [0,1]      \\[0pt]
                 & to the density of species $i$ &				\\
$\psi_i$ & Sensitivity of predation in species $i$    & [0.5,1.5]    	\\[0pt]
                & to the density of predators &				\\
     \hline
\end{tabular}
 \caption{Generalized model parameters as defined in Ref.~\onlinecite{gross_generalized_2006}. For numerical simulations, the turnover rates $\alpha_i$ are scaled by the species' fitness $n_i$ corresponding to the niche value used to generate the underlying topology \cite{williams_simple_2000}, while the other parameters are drawn from the indicated ranges.}
\label{tab:Parameters}
\end{table}

In summary, the identity \eqref{eq:Identity} allows one to break the partial derivative of the process in the steady state into two constant factors, describing the per-capita rate and the sensitivity of the process, respectively.
These factors are therefore well-defined ecological parameters in their own right, which can be understood and discussed even if the steady state of the system is unknown. For food webs, this parametrization leads to the Jacobian matrix expressions given in the online material \cite{gross_generalized_2006} and the parameters given in \tabref{tab:Parameters}.
Using \eqref{eq:ImpactMatrix}, the Jacobian that is thus parametrized can then be employed to relate a perturbation to its eventual impact. 
In the following, we use this approach to discuss the prediction of the impact of perturbations on food webs.

\subsection{Remarks}

We note that the approach to assessing impact taken in this paper is closely related to \cite{novak_predicting_2011}. Our main methodological contribution is to apply this approach to generalized models. The advantage of generalized modeling 
is its high numerical efficiency, which enables a detailed and statistically sound numerical exploration. 
For the practical application to real world food webs, generalized models offer additional advantages. 
In contrast to half-maximum concentrations and maximal growth rates used in conventional models, all parameters of the generalized model are defined in the state observed in nature. The parameters can therefore be measured directly without requiring a fitting procedure. Furthermore, the parameters are defined in such a way that their estimation from noisy data converges maximally fast \cite{houthakker_income_1969,lade_early_2012}. 

The formulation of the generalized model is straightforward (see the online material for a short review). Based on Ref.~\cite{gross_generalized_2006}, or with a parametrization algorithm \cite{aufderheide_cip_2013}, even large models with tens to hundreds of species can be set up in few hours. The impact of different species can then be computed in seconds on a small laptop, using a simple algorithm \cite{aufderheide_cip_2013}. Once the model has been set up, integration of new data requires entering new numerical values. The computation of impact and importance therefore presents only a small additional effort to the field work needed to measure parameters.

\begin{figure}[tb]
\includegraphics{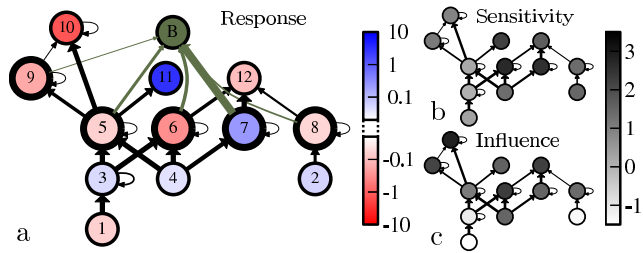}
 \caption{Predicted response of the Gatun Lake food web \cite{zaret_species_1973} to the perturbation caused by the introduction of the peacock bass. Shown are species (circles) and predator-prey relationships (arrows, in the direction of the biomass flow).
In panel a, the peacock bass population (B) is shown, together with its feeding relations. The colour code denotes the response of each species to this perturbation for a specific set of generalized parameter values (see the online material for details).
 In b) and c) the grey-scale denotes the sensitivity and the influence of each species respectively, approximating its average propensity to be impacted by other species or to impact to other species (details are given in \secref{sec:importance} and the online material). Colour online.}
\label{fig:ImpactWeb}
\end{figure}

\section{Examples \label{sec:examples}}
For illustration, we consider two examples, a simple predator-prey system and the real-world food web of Gatun Lake shown in \figref{fig:ImpactWeb}. While the predator-prey system is analytically tractable, the second example requires numerical calculations.

As the first example, we consider a class of predator-prey models in which a predator of abundance $X_1$ consumes a producer of abundance $X_2$. A detailed treatment and discussion of the stability of this system in terms of the generalized model parameters can be found in Refs.~\onlinecite{gross_enrichment_2004,gross_generalized_2006}. 

The Jacobian matrix of this system near the steady state is
\eq{\matr{J}=\begin{pmatrix}\alpha_1(\psi-\mu)&\alpha_1\gamma\\
			    -\alpha_2\sigma\psi &\alpha_2(\phi-\sigma\gamma-\tilde\sigma\mu),
             \end{pmatrix}}
where $\alpha_i$ represents each species' turnover rate, $\sigma$ the relative loss of the producer due to predation (instead of natural mortality), $\phi$ the elasticity (i.e.\ sensitivity) of primary production to the producer abundance, $\gamma$ the elasticity of predation to primary producer abundance, $\psi$ the elasticity of predation to predator abundance, and $\mu$ the elasticity of natural mortality to a species' own abundance. 

We now consider the impact of the arrival of a competing predator in the established producer-predator system.
It can be assumed that this new predator has a direct negative effect on the primary producer but no direct effect on the established predator, such that the perturbation matrix ${\matr{K}}$ contains the entries $K_{1,1}=0$ and $K_{2,1}<0$. 
As shown in detail in the online material, the impact on the established predator and producer are 
\eq{I_1=\frac{\alpha_1}{\det{\matr{J}}}\gamma K_2 \quad \text{and}\quad I_2=\frac{\alpha_1}{\det{\matr{J}}}(\mu-\psi) K_2,} where $\det\matr{J}\geq0$ represents the determinant of the Jacobian matrix. 

We see that generally the impact on the established predator is negative. This result is intuitive as the established predator is now in exploitative competition with the arriving predator.
Of particular interest is the case where the established predator suffers from linear loss ($\mu=1$) and has an effect on the producer that scales linearly with the predator abundance ($\psi=1$, i.e.\ there is no interference between the predators). In this limit, $\det \matr{J}$ approaches $\alpha_1\alpha_2\sigma\gamma$. For the impact on the producer $I_2=0$ because $\psi-\mu$. However, the impact  on the established predator, $I_1=\ttfrac{1}{(\alpha_2\sigma_2\gamma_1)}$, is large. This is a manifestation of the well-known competitive-exclusion principle, which precludes the coexistence of the predators in this case \cite{hardin_competitive_1960,gross_invisible_2009}. 

The assessment of impact in larger food webs can be carried out analogously, but  requires numerical computations in which the generalized parameters are set to specific values. Here, we illustrate this assessment by the example of the Gatun Lake food web \cite{zaret_species_1973} (\figref{fig:ImpactWeb}) for a simple set of such parameters.
Specifically, we focus on the impact of the population  of a predatory fish, such as the peacock bass \emph{Cichla ocellaris} (\figref{fig:ImpactWeb}a). A step-by-step breakdown of the parametrization and perturbation assessment is presented in the online material.  

The proposed approach predicts that peacock bass have a strong and generally negative impact on the secondary consumers on which they feed. Furthermore, they have a generally positive impact on the consumer's prey, and a generally negative impact on other top predators. These observations are consistent with basic ecological reasoning.
Counter-intuitive results are found for species 11 (black tern), which benefits from the strong  decrease of its competing predators 9 (bigmouth sleeper) and 10 (tarpon), and for species 7 (sailfin molly, mosquito fish), which benefits from the reduction of 6 (tetras) with whom they are both in exploitative and apparent competition.

Real world observations of the Gatun Lake showed \cite{zaret_species_1973}, that the introduction of the peacock bass strongly decreased  the secondary consumers (5--7) and their predators (9--12), but increased the consumers prey (3--4) and, counterintuitively, in species 8 (blackbelt cichlid). Comparing the predictions of the model with the real data, we note that the model correctly predicted the change in the producers 1--4, the decrease in the consumers 5 and 6, and the decrease in predators 9, 10, and 12. 

However, the model predictions do not agree with the observed decrease in species 11 (black tern), and 7 (sailfin molly, mosquito fish) and with the increase in species 8 (blackbelt cichlid). A likely reason for the discrepancy between predicted and observed results is the inaccuracy of the parametrization used here, that we based on simple allometric scaling relationships instead of direct observations. Reviewing the results, we can guess that the incorrect prediction for an increase in species 11 is due to an overestimation of the sensitivity of its competing top predators (9--10) to the decrease in their common prey 5 (silverside). Further, the incorrect prediction for an increase in 7 and a decrease in 8 might result from incorrectly estimating the benefit to the prey of 12 (herons and kingfishers) due to the strong decrease in 6 (tetras).  This exercise shows how food webs often follow the basic logic of direct and indirect effects, but that the details can be sensitive to incorrect parametrization.

\begin{figure}[tb]
\includegraphics{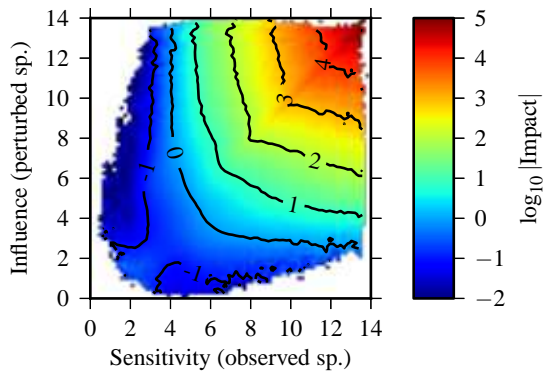}
\caption{
The average absolute impact on a species with given sensitivity if a species of given influence is perturbed. Generally, the more sensitive the species being tracked, and the more influential the species that is increasing, the greater the change (impact) to the observed species. Other Parameters $N=50, C=0.04$, see online supplement for $C=0.08$.
\label{fig:ImpoImpoImpact}}
\end{figure}

\section{Sensitive and Influential Species\label{sec:importance}}

To obtain better theoretical predictions for real world systems, such as Gatun Lake, a more precise parametrization of the model is highly desirable. However, not all species are of equal importance in the food web and thus also precise information is more important for certain parameters than for others. To gain insights into the importance of various species in the web, we now identify \emph{sensitive species} are easily perturbed by disturbances propagating through the web, and \emph{influential species} have a strong effect on other species, when directly perturbed. 

Close to a steady state, the dynamical properties of a system are characterized by its dynamical modes which consist of the eigenvectors and eigenvalues of the Jacobian matrix \cite{kuznetsov_elements_1998}. 
For a given matrix, there are generally two different sets of eigenvectors, which are called right and left eigenvectors \cite{wong_computational_1997}. For each eigenvalue $\lambda_k$ of the matrix there is thus a corresponding right eigenvector $\vect{v}^{(k)}$ and left eigenvector $\vect{w}^{(k)}$.

One can visualize dynamical modes as vibrations travelling through a drum when it is struck.
Here the different modes correspond to different notes that are played on the drum.
The right eigenvectors characterize the pattern of vibration when a specific note is played. 
Specifically, the elements of the right eigenvector describe how strongly the respective area of the drum vibrates in that note.
The same is true for the food web. 
In a stable steady state that is hit by a short (pulse) perturbation the right eigenvectors govern how the system returns to the steady state after the perturbation. 

Drummers know how to play different notes by striking different parts of the drum. 
This is captured by left eigenvectors. 
Specifically, the elements of the left eigenvector for a given dynamical mode describe 
how strongly the specific mode is excited when the drum is struck in a given area.   
Similarly, in the food web, the left eigenvectors characterize the strength of a specific dynamical response when a given species is perturbed. 

Intuitively, one can think of each dynamical mode as a possible response of the system to a perturbation. More precisely, the right eigenvector denotes the impact of response (which species ``feel the vibrations''), while the corresponding left eigenvector denotes the type of perturbation that can trigger a particular response (which species needs to be perturbed to ``play a given note''). For instance, consider the pair of a right eigenvector $\vect{v}=(1,2)$ and a left eigenvector $\vect{w}=(1,0)$. This mode can only be excited through perturbation of the first species, but when it is excited the second species feels the system's response twice as strong as the first.

The strength of a mode's response is determined by its \emph{excitability}. 
The excitability is, for a given mode, given by the corresponding eigenvalue $-\ttfrac{1}{\lambda_k}$ of the inverse Jacobian matrix $\matr{J}^{-1}$, which has the same eigenvectors as $\matr{J}$. Intuitively, the (negative) eigenvalue $\lambda_k$ of a dynamical mode indicates a system's resistance to a particular perturbation. The impact of such a perturbation is therefore inversely proportional to this resistance  (see the online material for more details). 

In the case presented here, we consider that a perturbation continuously excites the same dynamical modes (press perturbation). The impact is therefore the combined continuous excitation resulting from the perturbation of these dynamical modes.

The potential impact that a species feels due to a given dynamical mode is the product of the mode's excitability and the component of the right eigenvector on this species. Furthermore, the potential impact from all modes is the sum over the contributions from the individual modes $k$. Taking the logarithm of this sum to bring the numerical values into a more manageable range, we therefore define the \emph{sensitivity} of a species, $\mathrm{Se}_i=\log(-\sum_k\ttfrac{\ttabs{v^{(k)}_i}}{{\lambda_k}})$, where $\ttabs{v^{(k)}_i}$ is the absolute value of the entry $v_i$ of the right eigenvector corresponding to mode $k$. For a more formal derivation, see the online material.

The potential impact that a species causes by exciting a given dynamical mode is the product of the mode's excitability and the component of the left eigenvector on this species.
Analogously to the sensitivity, we therefore define the \emph{influence}, $\mathrm{In}_i=\log(-\sum_k\ttfrac{\ttabs{w^{(k)}_i}}{{\lambda_k}})$.

\begin{figure}[tb]
\includegraphics{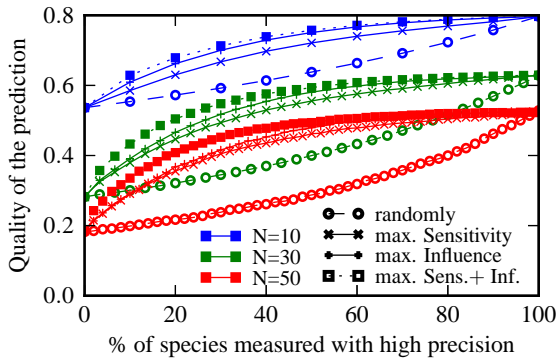}
\caption{Focusing on sensitive and important species can reduce the measurement effort when species are measured successively with higher precision. Starting on the left, only low-precision information is available for all food web species. Advancing to the right, the measurement error is reduced for one species at a time until all nodes have been measured with high precision. The different curves refer to different strategies for selecting which species to measure precisely. For the dashed line (empty symbols), the species are selected randomly. For the solid curves, we first evaluate the sensitivity or influence  of each species based on current knowledge and then select the species with the highest value  of either sensitivity or influence. For the dotted curves we select the species with the highest product of influence and sensitivity ($\mathrm{Se}_i\mathrm{In}_i$). Error-carrying parameters of each node: $\alpha,\mu,\psi,\phi,\gamma,\sigma,\beta,\chi$. 
Other Parameters: Initial error of each parameter $10\%$, final error $2\%$, connectance $C=0.04$. Higher connectance (see online material) results in similar graphs, but the overall prediction quality decreases.}
\label{fig:NumExp}
\end{figure}

The example of the Gatun Lake food web  shows that the general results for influence and sensitivity are largely consistent with Gatun Lake's response to the introduction of peacock bass (\figref{fig:ImpactWeb}). Because the perturbation affects several influential nodes (e.g. species 6 and 9), most of the species of the food web are affected. In particular sensitive nodes (e.g. species 6, 7, 11) respond strongly to the perturbation (see online material for more details).

To further confirm the relationship between importance, sensitivity, and actual impact of a perturbation, we consider a ensemble of $10^6$ randomly generated food webs with $50$ species and average connectance $0.04$ generated as described in Ref.~\onlinecite{gross_generalized_2009} and as reviewed in the online material. The topology of these food webs is generated using the niche model \cite{williams_simple_2000}. For the numerical computations, the generalized parameters of the species in this topology are then drawn uniformly and independently from the ranges given in \tabref{tab:Parameters}. For these food webs the colour code in \figref{fig:ImpoImpoImpact} indicates the average impact that a focal species of given sensitivity experiences when a species of given importance is perturbed. This reveals a strong correlation of the impact with both the sensitivity of the focal species and the importance of the perturbed species. 

Considering the ensemble of model food webs again, we observe that the number of the very influential and very sensitive species in each food web is small. For instance, we find that on average for each web only $15.4\%$ of all species have a sensitivity value in the upper $30\%$ of the sensitivity range for this web, and only $18.6\%$ have an importance value in the upper $30\%$ of the influence range for that web.

Summarizing the above, knowledge of the Jacobian of a specific food web enables us to predict the impact of specific perturbations, and also allows us to gain a more general understanding of the species' sensitivity and influence with regard to perturbations of the network. The main challenge for impact assessment is thus to collect the necessary data for constructing the system's Jacobian. As hinted previously, and shown below, precise measurement is required only for some species.

\section{Iterative Parameter Estimation\label{sec:errors}}
It is intuitive to assume that accurate predictions hinge on precise measurements of the parameters of the most influential and sensitive species.
However, our approach to identifying these species builds on analysis of the Jacobian and thus itself requires the same type of information. It is thus not possibly to determine a-priori which species are important or influential. 
To address this dilemma, we now propose an iterative strategy in which existing preliminary information is used to estimate the impact and sensitivity of species.
This assessment is then used to obtain improved parameter estimates on seemingly important species. 
Once additional data on these species becomes available, they can be used to further improve the estimates of the impact and sensitivity of species, refining the process. 
Thus, a cycle is formed in which the necessary information for precise impact predictions is iteratively assembled. 

We explore the quality of impact prediction in a series of numerical experiments.  
In each experiment, the task is to predict the impact of a random perturbation to a food web that is generated according to the procedure described in \secref{sec:importance}. This procedure determines values of the generalized parameters of the \emph{true} Jacobian of the food web by drawing them uniformly from the ranges indicated in \tabref{tab:Parameters}. Based on this true Jacobian, we additionally generate an \emph{estimated} Jacobian with slightly different generalized parameter values to simulate measurement errors. More precisely, we draw each generalized parameter value used in the estimated Jacobian from a log-normal distribution centred on the corresponding parameter value used in the true Jacobian; the log-normal distribution is chosen to allow large errors while keeping the sign of parameters consistent. We then compute the true impact of the random perturbation, $\vect{I}$ based on the true Jacobian, and the estimated impact $\tilde{\vect{I}}$ based on the estimated Jacobian.
The quality $Q$ of the impact estimation is then evaluated as the cosine of the angle between the true and the estimated impact vectors (see online material for details).

Now, we introduce a numerical implementation of the iterative strategy described above. 
We consider numerical experiments in which the knowledge of the Jacobian is initially  poor, such that the generalized parameters are drawn from a lognormal distribution with a standard deviation of $10\%$ of the true value. We furthermore assume that additional empirical work can be carried out on specific species that brings the error in all parameters of the respective species down to $2\%$. 
Our aim is to carry out the precise measurements in the order that leads to the most rapid increase in the quality of impact prediction. 

For the purpose of demonstration, we consider four different protocols: a) precise measurements are carried out in random order, b) species are measured in the order of decreasing influence, c) species are measured in the order of decreasing sensitivity, d) species are measured in the order of the decreasing sum of sensitivity and influence. The choice of species to measure next, was always based on the \emph{estimated} Jacobian that is available at the time. Thus, only information is used that would also be available in the real world at the respective time. 

\begin{figure*}[tb]
\includegraphics{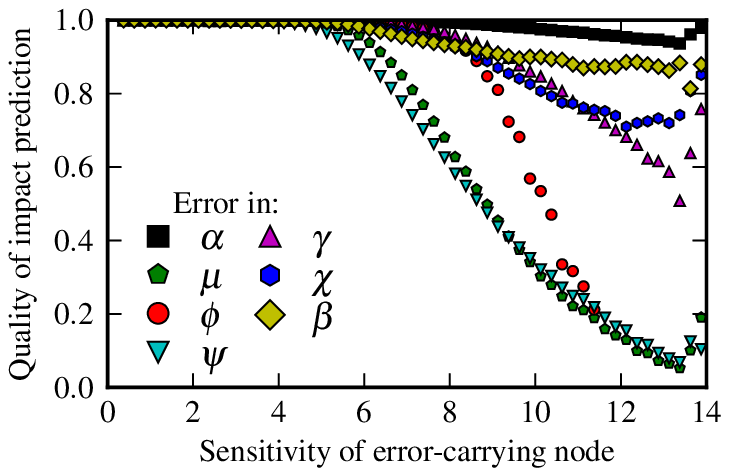}
\includegraphics{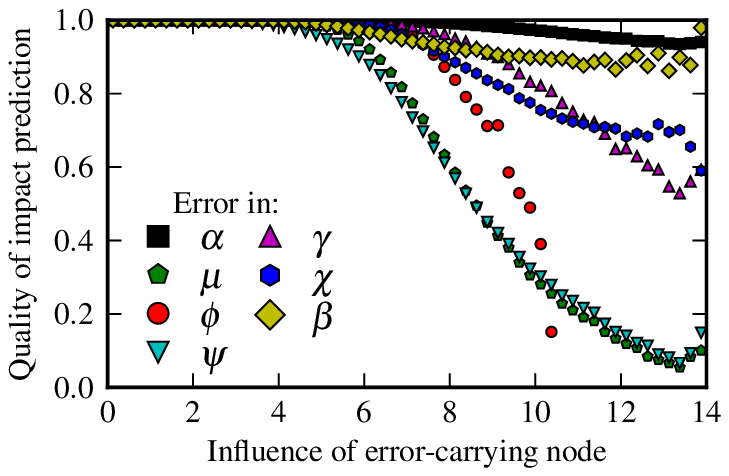}
\caption{Average quality of an impact prediction in the presence of measurement errors. The average quality of an impact prediction if one node with the specified sensitivity (left) or influence (right) in a food web is subject to a measurement error. The different datasets refer to errors in different parameters (c.f.\ \tabref{tab:Parameters}). Other parameters are: system size $N=50$, connectance $C=0.04$, and the standard deviation of the relative measurement error $10\%$. For higher $C$, the effect of $\chi$ and $\beta$ increases (see online material). Noise at high importance and sensitivity values is due to the relatively rare occurrence of these values in the numerical experiments. 
\label{fig:ImpactSensi}}
\end{figure*}

The results shown in \figref{fig:NumExp} demonstrate that estimating influence and sensitivity of the species prior to each measurement strongly increases the accuracy of predictions. This is particularly pronounced if measurements focus on the species with the highest product of sensitivity and influence. For instance, after measuring $20\%$ of all species according to this protocol, we attain a quality of prediction comparable to measuring $60\%-80\%$ of all species when species are chosen randomly. Using the estimation of influence and sensitivity to focus observational or experimental efforts can thus significantly reduce the amount of empirical work that is needed to achieve a given prediction quality.

\section{Most Important Parameters and Species to Measure\label{sec:statistical}}

The iterative refinement procedure proposed above, needs some initial information on the system as a starting point.in this section, we therefore explore what types of parameters and what types of species were most important to measure.

To get an initial intuition of the importance of different parameters for impact prediction, we consider a situation where the estimated Jacobian is identical to the true Jacobian except for a single parameter that carries an error. 
The quality of the estimated impact decreases with increasing influence and sensitivity of the species affected independently from the varied parameter (\figref{fig:ImpactSensi}).
Furthermore, \figref{fig:ImpactSensi} shows that the decrease in quality for sensitive and influential species depends on the parameter under consideration; precision in the elasticity of the mortality $\mu$, and of the elasticity of predation $\psi$ with respect to predator abundance were the most important.
This confirms our intuition that not all parameters need to be measured to the same level of accuracy.

To determine which species are most important to measure in the absence of knowledge about the Jacobian, we explore the correlations between sensitivity or influence and species properties (see below) in a set of $10^6$ model food webs.
In the analysis, we consider the correlations (not causal effects) of sensitivity and influence with the following potential biological indicators:   
\begin{itemize}
\item Generality, or the number of prey species of the focal species.
\item Vulnerability, or the number of predators of the focal species.
\item A binary value that is 1 if the focal species is a primary producer and 0 otherwise.
\item The trophic level $TL$, which we calculate by solving a set of linear equations, such that  $TL_i=1$ for primary producers and $TL_i=1+\text{mean}(TL_{prey})$ for consumers, where $\text{mean}(TL_{prey})$ denotes the mean trophic level of $i$'s prey. 
\item The biomass turnover rate (generalized model parameter $\alpha$), indicating the amount of biomass an individual consumes in comparison to its own mass.
\item The weighted topological importance of a species  $WI^s$ as introduced in detail in Ref.~\onlinecite{jordan_topological_2006}. For each species, the value of $WI^s$ indicates the indirect interactions from other species, based on the topology and biomass flows of a food web. The step parameter $s$ indicates the maximum number of direct interactions, through which indirect effects are perceived. 
\end{itemize}
If for a given web, $x_i$ indicates one of these properties for species $i$ in this web, and $y_i$ its sensitivity or influence. Then we denoted the (Pearson) correlation coefficient between $x$ and $y$ as $R=\ttfrac{\overline{(x_i-\bar{x})(y_i -\bar y)}}{(s_x s_y)},$ where $\bar{x}$ denotes the mean of $x_i$ over all species $i$, and where $s_x$ denotes the standard deviation of the $x_i$.

\begin{figure}[tb]
\includegraphics{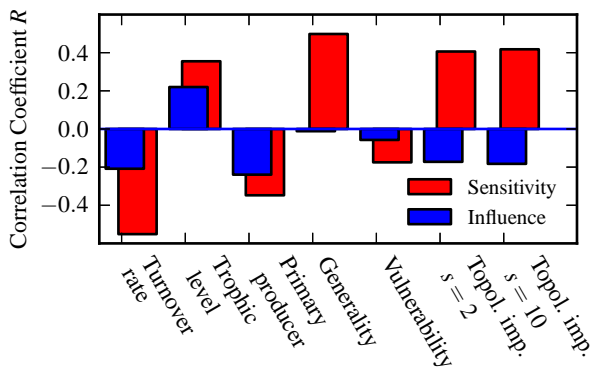}
\caption{
Correlations of species' properties with their sensitivity and influence. The pattern of correlations is consistent  with large predators playing an important role for a system's response to perturbations. Parameters $N=50, C=0.04$, results are similar for higher connectance (see online supplement).}
\label{fig:ImportanceProps}
\end{figure}

The correlation analysis (\figref{fig:ImportanceProps}) shows that high trophic levels and low biomass turnover rates (e.g., long life span) correlate strongly with sensitivity and influence. This result suggests that top predators and other large and long-lived species, despite their typically small total biomass, play a disproportionate role in the systems' response to perturbations. This corresponds well with the observation in real-world systems \cite{zaret_species_1973,berg_using_2011,moyle_fish_1996}. When no specific information on biomass flows is available, these species should be targeted for initial parametrization.

The sensitivity of a species is highly correlated with its generality, while its influence appears to be independent of its generality.  Intuitively, this can be interpreted as generalist species being sensitive to all of their prey species, while having relatively little impact on those species.

The weighted topological importance correlates strongly with sensitivity and weakly with influence.     
This suggests that indirect effects from proximal species in the web play generally a large role in the response, while indirect effects from distal species are relatively minor.

To improve perturbation assessments in the Gatun Lake ecosystem, field work should prioritize better measurements of species $6$ (tetras), $7$ (sailfin molly, mosquito fish), and, to a lesser degree, the large predators 10--12 that have the highest combined sensitivity and influence (\figref{fig:ImpactWeb}). Even, if the explicit sensitivity and influence values were not available, the correlation would suggest to prioritize these species, because of their low turnover rates, relatively high generality and their high trophic levels. Furthermore, the most significant improvements should be obtained by focusing in the measurement process on the elasticities of mortality $\mu$ and of predation $\psi$ of these species.

\section{Conclusions}
Previous work has suggested that without near perfect information on a range of parameters, it may be
intractable to predict the effects of perturbations to large, complex systems \cite{novak_predicting_2011}.
In this paper, we proposed a method to predict the impact of perturbations on complex systems more efficiently.
We used this method to investigate the relative importance of different species in food webs.  
We found that there are typically a small number of species that are highly important, because 
they react sensitively to perturbations, have a strong influence on others, or both. 

The proposed method is based on the linear stability of steady states in food webs. Strictly speaking, it therefore  describes the consequences of small perturbations to systems close to a state of stable species coexistence. However, linear stability is often found to agree well with other stability criteria for food webs, such as robustness against noise \cite{pimm_food_2002}, or permanence which measures the boundedness of a trajectory in a plausible part of the state space \cite{chen_global_2001}. One can therefore expect that our methods give at least some indications about perturbation consequences and key species in  systems that are not in a steady state. 

While we have focused exclusively on food webs, we note that the same approach can likewise be applied to other networks of nonlinear interactions that are found in metabolism \cite{steuer_structural_2006}, gene regulation \cite{gehrmann_boolean_2010}, and cellular population dynamics \cite{zumsande_general_2011}.

The presented results suggest that the potential impact of environmental perturbations on food webs 
can be predicted with reasonable accuracy if the most relevant parameters for only a small number of important species in the web are measured well.

Finding the important species in a food web appears as the key to a good and efficient impact assessment.
We propose to find these important species by a) pre-selecting species based on their biological properties and b) applying an iterative refinement method once some initial information is available. 

Our correlation analysis suggests that it is most important to obtain precise parameter estimates for long-lived, generalist consumers at high trophic levels. Furthermore, our analysis suggests that for these species, it is most important to precisely estimate the dependence of their mortality and predation on their abundance.

\section*{Acknowledgments}
We thank Prof. Mark Novak for discussions.

\addcontentsline{toc}{section}{Bibliography}
\bibliographystyle{unsrt}

\end{document}